# Universal Organization of Resting Brain Activity at the Thermodynamic Critical Point

Shan Yu,[1,§] Hongdian Yang,[1,§] Oren Shriki,[1] Dietmar Plenz[1,*]

[1] Section on Critical Brain Dynamics,
National Institute of Mental Health, NIH, Bethesda, MD, USA

[§] These authors contributed equally to this work.
[*] Correspondence should be addressed to plenzd@mail.nih.gov

**Abstract:** Thermodynamic criticality describes emergent phenomena in a wide variety of complex systems. In the mammalian brain, the complex dynamics that spontaneously emerge from neuronal interactions have been characterized as neuronal avalanches, a form of critical branching dynamics. Here, we show that neuronal avalanches also reflect that the brain dynamics are organized close to a thermodynamic critical point. We recorded spontaneous cortical activity in monkeys and humans at rest using high-density intracranial microelectrode arrays and magnetoencephalography, respectively. By numerically changing a control parameter equivalent to thermodynamic temperature, we observed typical critical behavior in cortical activities near the actual physiological condition, including the phase transition of an order parameter, as well as the divergence of susceptibility and specific heat. Finite-size scaling of these quantities allowed us to derive robust critical exponents highly consistent across monkey and humans that uncover a distinct, yet universal organization of brain dynamics.

## Introduction

The cerebral cortex of the mammalian brain consists of tens of billions of neurons and interactions among them exist at various scales ranging from local microcircuits, to cortical areas, and even across the entire cortex. This myriad of neuronal interactions underlies various brain functions including motion, perception and cognition [1–4]. Understanding the general principles governing neuronal interactions and how they give rise to emergent properties of information processing is among one of the most challenging questions in systems neuroscience.

For several decades, concepts and tools developed in statistical physics address the collective behavior of complex systems by studying the interactions among the constituent microscopic system components. Of the many states a complex system might adopt, the critical state at thermodynamic equilibrium has been extensively studied and this state might be particularly relevant for the brain. Microscopically, the critical state represents exquisitely balanced interactions among all system components [5]. Macroscopically, such balanced interactions poise the system at a transition between two contrasting phases (quantified by the order parameter, $M$) and give rise to a number of non-trivial emergent properties, including the divergence of the sensitivity to external perturbations (quantified by the susceptibility, $\chi$), and internal complexity/diversity (quantified by the specific heat, $C$) [6–8]. For a cortical system, these quantities have intuitive meanings in terms of neuronal information processing. $\chi$ reflects the input sensitivity of the system [9], $C$ reflects the dynamic range of neuronal populations in representing inputs [10,11], and $M$ measures the overall activity level in the cortical network. The maximization of $\chi$ and $C$ brought by a critical state can thus be seen as optimizing input sensitivity [12–14] and dynamic range [10,11], respectively. At the same time, the phase transition of $M$ (overall activity level), may reflect state changes in the brain, such as sleep/awake and inattentive/attentive transitions [15–19].

Importantly, near the critical state, those emergent behaviors do not depend on the specific microscopic realization of a system. It has been shown that a multitude of systems can be categorized into a small number of "universality classes" based on only a few parameters, i.e., so called "critical exponents" [5–8]. Within individual classes, apparently different systems follow the same quantitative rules. A major question thus arises, whether such universality of critical behavior, encountered when studying physical systems, might also include biological complex systems such as the cortex that evolved to process information.

Recent studies of neuronal avalanches strongly suggest that neuronal interactions, both at the mesoscopic scale [20,21] (within tens of $mm^2$ of cortical tissue) as well as macroscopic level [22–25] (across the entire cortex), may position the cortex at or near a non-equilibrium critical state in order to optimize information processing [14,26-29]. Neuronal avalanches are intermittent cortical activity cascades that spontaneously form in the normal brain. During an avalanche, spontaneous activation of one neuronal group can trigger consecutive activations of other neuronal groups within just a few milliseconds and the propagation of such activity spans both spatial and temporal domains. This propagation is well described by a non-equilibrium critical branching process, which successfully explains some of the functional advantages of neuronal avalanches [14,20,27,29]. However, it is currently unclear if neuronal avalanches indicate cortical dynamics close to a critical state in the equilibrium thermodynamic sense and, if so, what universality class the cortical activities may belong to. The current study is aimed to address these questions and their potential functional implications for the brain.

## Materials and Methods

**Local field potential (LFP) and Magnetoencephalography (MEG) recordings**

Ongoing LFP activity was recorded from two adult monkeys (*macaca mulatta*). Multi–electrode arrays (10 ×10, 400 *μm* inter–electrode distance, 1 or 0.6 *mm* electrode length, BlackRock Microsystems) were chronically implanted in the left pre–motor (Monkey 1) or prefrontal (Monkey 2) cortex (Fig. 1A). 20-30 min of ongoing LFP (1-100 Hz) signals were simultaneously obtained from each electrode while the animals were sitting alert in a primate chair but not engaged in any behavioral task. For more experimental details, see [30]. Ongoing MEG activity (~30 min, 1-150 Hz) was recorded from 3 healthy human participants (female). The sampling rate was 600 Hz. The sensor array consisted of 273 axial first-order gradiometers. Analysis was performed on the axial gradiometer waveforms. For more details, see [22].

**Avalanche analysis**

Negative deflections in the LFP (nLFPs) were detected by applying a threshold at -2.5 standard deviations (SD) of the LFP fluctuations estimated for each electrode separately (Fig. 1B). Such a threshold is based on the non-linear relation between nLFP amplitudes and ability of local neuronal groups to synchronize with other, spatially separated ones [30,31]. The nLFP peak times were then binned using a time window, $\Delta t$. Results shown are based on $\Delta t = 2$ *ms* (Monkey 1) and 4 *ms* (Monkey 2) but they are similar across a wide range of $\Delta t$ (2-16 ms tested). Spatiotemporal clusters of nLFPs, i.e., avalanches, were defined by consecutive bins such that each bin contained at least one nLFP at any site in the selected group [20]. The size of a cluster, *s*, was defined as the number of nLFPs in the cluster (Fig. 1C). Similar analysis was applied to identify avalanches from the MEG recordings, for which a threshold at -3.0 SD of the MEG waveforms was used to detect significant neuronal events. The time window, $\Delta t$ was 1.67 (1×sampling period; subject 1) or 3.34 *ms* (2×sampling period; subjects 2, 3). For more details, see [22]. Avalanche patterns were obtained by collapsing all time bins within an avalanche to form a corresponding spatial pattern $\boldsymbol{\sigma} = (\sigma_1, \sigma_2, ..., \sigma_n)$, where *n* is the number of recording sites, i.e., system size, included in the analysis and $\sigma_i = 1$ if at least one nLFP occurred at site *i* and $\sigma_i = -1$ otherwise (Fig. 1C).

**Using the dichotomized Gaussian (DG) model for estimating pattern probabilities $p_i$**

The DG model is a useful tool for capturing the statistics of binary neural activity patterns [11,30,32,33]. It applies a thresholding operation to multivariate Gaussian variables: $y_i = 1$ when $u_i > 0$ and $y_i = -1$ otherwise, where $\mathbf{u} = (u_1, u_2, ..., u_n) \sim N(\delta, \lambda)$, $\delta$ is the mean and $\lambda$ is the covariance of the Gaussian variables. In order to match the rate, *r*, and covariance, $\Sigma$, of the observed binary variables, i.e., avalanche patterns, $\delta$ and $\lambda$ need to be adjusted according to $\delta_i = \Phi^{-1}(r_i)$ and $\lambda_{ij}$ as the solution for $\Sigma_{ij} = \Phi_2(\delta_i, \delta_j, \lambda_{ij}) - \Phi(\delta_i) \Phi(\delta_j)$, where $\Phi$ and $\Phi^{-1}$ are the cumulative probability function of a Gaussian distribution ($\Phi$ for 1-dimensional and $\Phi_2$ for 2-dimensional) and its inverse function, respectively. An implementation of the model in MATLAB can be found in [33]. The pattern probabilities for the DG model were obtained by calculating the cumulative distribution of multivariate Gaussians (MATLAB function *mvncdf*).

**Fitting a power-law to the size distribution**

The exponent of the best fitting power-law, was estimated by minimizing the Kolmogorov-Smirnov (*KS*) distance between the empirical distribution and a power-law distribution [34]. The KS distance ($D_{KS}$) was defined as

$$D_{KS} = \max_s | CDF_{emp}(s) - CDF_{power-law}(s) | , \qquad (1)$$

where *s* is the pattern size and $CDF_{emp}$ and $CDF_{power-law}$ represent the cumulative distribution function for the empirical size distribution and the power-law function used for fitting, respectively.

**Inferring $p_i$ for different values of *T***

To predict the pattern probabilities $p_i$ for different values of the fictitious temperature, *T*, it is useful to express the state probability as a function of interactions that occur at different orders [32,35]. Let the pattern probability be $p(\boldsymbol{\sigma})$, where $\boldsymbol{\sigma} = (\sigma_1, \sigma_2, ..., \sigma_n)$ and $\sigma_n = \{1,-1\}$, representing the states of individual components. Generally, we can write $p(\boldsymbol{\sigma})$, using the full log-linear expansion, as

$$p(\boldsymbol{\sigma}) = \frac{1}{Z} \exp\left( \sum_i \theta_i \sigma_i + \sum_{(i<j)} \theta_{ij} \sigma_i \sigma_j + \sum_{(i<j<k)} \theta_{ijk} \sigma_i \sigma_j \sigma_k + ... \right), \qquad (2)$$

where *Z* is the normalization factor and $\theta$ characterizes different orders of interactions. The full log-linear expansion and its lower-order approximations have been widely used in characterizing neuronal interactions [36-38].

Next, we define $\theta = \theta_0/T$, where $\theta_0$ represent the intrinsic interaction strength that does not depend on *T*. If we denote $E(\boldsymbol{\sigma}) = -\left( \sum_i \theta^0_i \sigma_i + \sum_{(i<j)} \theta^0_{ij} \sigma_i \sigma_j + \sum_{(i<j<k)} \theta^0_{ijk} \sigma_i \sigma_j \sigma_k + ... \right)$, Eq. 2 can be rewritten as

$$p(\boldsymbol{\sigma}) = \frac{1}{Z} \exp\left( \frac{-E(\boldsymbol{\sigma})}{T} \right). \qquad (3)$$

We can then use the single histogram method [9,39] to infer $p_i$ for different *T*, an approach that was used for modeling natural image statistics [40] and was also recently introduced to neuroscience [10]. Specifically, if $p_i$ denotes the probability of any given pattern *i* and $E_i$ the corresponding *E*, Eq. 3 changes to

$$p_i = \frac{1}{Z} e^{-E_i/T} . \qquad (4)$$

Setting *T*=1 for the original recording, Eq. 4 can be expressed as

$$p_i(1) = \frac{1}{Z(1)} e^{-E_i}, \tag{5}$$

which enables us to compute $p_i$ for different $T$ as

$$p_i(T) = \frac{1}{Z} e^{\frac{-E_i}{T}} = \frac{1}{Z}[Z(1)p_i(1)]^{\frac{1}{T}} = \frac{Z(1)^{1/T}}{Z} p_i(1)^{1/T} \tag{6}$$

The normalization factor is determined by considering $\sum p_i(T) = 1$.

**Computing the specific heat, susceptibility, and order parameter**

The specific heat, $C$, is:

$$C = \frac{1}{n}\frac{\partial U}{\partial T} = \frac{\langle E_i^2 \rangle - \langle E_i \rangle^2}{nT^2}, \tag{7}$$

where $n$ is system size, $U \equiv \langle E_i \rangle = \sum p_i E_i$ and $E_i$ can be calculated according to Eq. 4. Given $n$ and $T$, $C$ reflects the variance of log ($p_i$), a useful metric for quantifying the capacity of the system to represent information [10,11].

The order parameter, $M$, is defined as:

$$M = \frac{1}{n}\sum_{i=1}^{2^n} p_i m_i, \tag{8}$$

where $m_i = \sum_{j=1}^{n} \sigma_j^i$. $\sigma^i$ indicates that the value of $\sigma$ is taken from the $i^{th}$ pattern. $M$ has a very intuitive meaning for a cortical system—it reflects the overall activity level of the system.

Finally, the susceptibility $\chi$ is a measure of the sensitivity of the system to small external perturbations. $\chi$ is defined as the change rate of $M$ when a small external field $H$ is applied:

$$\chi = \frac{\partial M}{\partial H}\bigg|_{H=0} = \frac{\langle m_i^2 \rangle - \langle m_i \rangle^2}{nT} \tag{9}$$

The field $H$ exerts its effect by changing the preference of the units to be active or not, i.e., their likeliness to be involved in an avalanche. Specifically, applying $H$ is equivalent to adding a term of $H\Sigma\sigma_i$ to the Hamitonian ($E$). For cortical dynamics, $H$ can be thought of as an approximation of a local perturbation, e.g., making a single or small group of neurons to fire (analog to flipping a single spin in a model; see [9] and/or a weak common input from, e.g., distant cortical areas or sub-cortical brain structures).

*Finite size scaling (FSS) analysis*

At the thermodynamic limit ($n \to \infty$), a critical system can be identified by power-law behaviors of its macroscopic quantities, including the correlation length $\xi$ (a characteristic distance beyond which correlations diminish), specific heat $C$, magnetization $M$ and susceptibility $\chi$. These quantities follow a power-law relation as a control parameter, such as the thermodynamic temperature $T$, approaches a critical value $T_c$, with specific critical exponents $\nu$, $\alpha$, $\beta$ and $\gamma$, respectively:

$$\xi \sim |t|^{-\nu} \tag{10}$$

$$C \sim |t|^{-\alpha} \tag{11}$$

$$M \sim |t|^{\beta} \tag{12}$$

$$\chi \sim |t|^{-\gamma} \tag{13}$$

where $t = (T - T_c)/T_c$. In principle, one could directly measure these relations to determine whether and when the system will be critical, i.e., to determine $T_c$, and, at the same time, estimate all critical exponents.

The complication comes with the fact that real systems are finite in size. This so called "finite size effect" causes the systems' behavior to deviate from the thermodynamic limit. Finite size scaling (FSS) is a standard procedure in statistical physics to solve this problem [7,9]. By analyzing the behaviors of systems with different sizes, FSS is able to extrapolate the results for the thermodynamic limit and correctly estimate $T_c$ and critical exponents. Briefly, we can choose a unique set of critical exponents to scale Eqs. 10-13 with different linear sizes of the system $L = \sqrt[d]{n}$, where $d$ is the dimensionality, and then collapse the curves obtained for all sizes. Specifically, $t$ needs to be scaled by $L^{1/\nu}$. Meanwhile, $C$, $M$ and $\chi$ are scaled by $L^{-\alpha/\nu}$, $L^{\beta/\nu}$ and $L^{-\gamma/\nu}$, respectively. The critical exponents ($\nu$, $\alpha$, $\beta$ and $\gamma$) and $T_c$ that can achieve the collapse is equivalent to a measurement made at the thermodynamic limit (see Supplementary Methods for details). Best collapse was achieved by minimizing the distance among all functions with different sizes using numerical optimization (MATLAB function *fminsearch*). Initial conditions for optimization are systematically changed with a grid search method in a wide range of parameter space and the resulting values for exponents were stable. These values are also stable with different ranges of *T* to perform FSS (the reported results were based on *T*=0.5-2.5).

**Measuring goodness of collapse**

For functions, e.g., $\chi_i$, corresponding to various system sizes *i*, the quality of the collapse is quantified by the ratio of mean squared deviation (MSD) for all functions to their average after the collapse to that before it. Formally, $MSD = \left\langle \left\langle (\chi_i - \bar{\chi})^2 \right\rangle_T \right\rangle_i$, where $\bar{\chi}$ is the point-wise average across system sizes, $\langle \ \rangle_T$ indicates the average across the range of *T* and $\langle \ \rangle_i$ indicates the average across system sizes. The closer the ratio to 0, the better the goodness of collapse.

# Results

**Avalanche dynamics at the mesoscopic scale**

We first investigated neuronal avalanches at the mesoscopic scale [20,21,30,41,42]. Ongoing neuronal activity in two monkeys was recorded with 10 ×10 high-density micro-electrode arrays chronically implanted in the superficial layers of cortex (Fig. 1A). Significant negative local field potential deflections (nLFPs), which indicate synchronized activity of local neuronal populations [21,30], were detected using an amplitude threshold of -2.5 standard deviations (SD) of the LFP calculated for each electrode (Fig. 1B). A spatiotemporal nLFP cluster was identified if nLFPs on the multielectrode array occurred within the same or consecutive time bins of width *Δt* (Fig. 1C). Importantly, the cluster size *s*, defined as the number of nLFPs in a cluster, distributed according to a power-law with an exponent close to -1.5. Moreover, the distribution exhibited scale-free behavior, i.e. the power-law and its slope were stable for different system size *n*, whereas the cut-off changed systematically with *n* (Fig. 1D-E). This power-law demonstrates that ongoing cortical activity at rest in awake monkeys organizes as neuronal avalanches [20,21]. It indicates the presence of significant correlations in neuronal activity among cortical sites and, accordingly, is destroyed when the times of nLFPs are shuffled randomly (Fig. 1D-E, broken lines).

**Characterization of the critical behavior**

Next we investigated whether neuronal avalanches reflect a cortical state close to thermodynamic/equilibrium criticality. Our general approach is based on a method similar to Monte Carlo simulations [9]. First, we estimate the probability $p_i$ of individual configurations in the system based on actual recordings, which gives a complete characterization of the system's behavior. Then, we infer the changes of $p_i$ with the change of a control parameter, $T$, which is equivalent to thermodynamic temperature. Finally, we compute various macroscopic properties including susceptibility, specific heat, and an order parameter, as a function of $T$ to judge if the actual $T$ (the one associated with the original recording) is close to the critical point.

More specifically, we define the configurations/states of the system by the spatial avalanche patterns, which were obtained by collapsing the spatiotemporal avalanche patterns along the temporal domain. This mapping ignores the internal temporal structure of individual avalanches. Each avalanche is originally represented by an $n$ by $m$ activity matrix, where $n$ is the number of electrodes and $m$ is the temporal extent of the avalanche. The activity matrix is then turned into an $n$-component binary vector where an electrode is set to 1 if it participates at least once in the avalanche and to -1 otherwise (Fig. 1C, see also Methods and [30]). The finite duration of the recording limits the direct estimation of pattern probabilities $p_i$ to $n \sim 10$. Therefore, in order to estimate $p_i$ for larger $n$, we take advantage of a simple parametric model, the Dichotomized Gaussian (DG) model [11,30,32,33], which considers only the observed first-order (event rate) and second-order (pair-wise correlations) statistics. This model estimates $p_i$ of avalanche patterns more accurately than directly measuring it from the limited data (supplementary Fig. S1; see also [30]). Due to the exponential increase in possible configurations with increasing $n$, we restrict the calculation of $p_i$ to $n \leq 20$. In total, we analyzed four 20-electrode sub-groups recorded from each of the two monkeys.

After obtaining $p_i$ for the condition in which the actual recording was taken, we introduce a control parameter $T$. $T$ is similar to the thermodynamic temperature, controlling both the likelihood of a given site to participate in an avalanche and the correlation among activities between different sites[7,9]. This allows us to systematically investigate the system's behavior at conditions different from the normal, physiological one. To infer $p_i$ for different $T$, we use the single histogram method [9,39], which accurately predicts behavior of equilibrium system for different values of the control parameter (see Supplementary Methods and Figs. S2 and S3 for validation of the equilibrium assumption). If we set the $T$ at which the actual recording was taken to be 1, it can be shown that $p_i(T) = \frac{1}{Z} p_i(1)^{1/T}$, where $p_i(T)$ is the state probability with the fictitious temperature $T$ and $Z$ is a normalization factor (Methods). After obtaining $p_i$ for a wide range of $T$, we use finite size scaling analysis (FSS) [9] to investigate whether the avalanche state ($T = 1$), is close to a thermodynamic critical point, i.e., if the critical "temperature" $T_c \approx 1$. We first analyze the thermodynamic quantities $\chi$, $C$, and $M$ as functions of $T$ for different system sizes ($n$=12-20; Fig. 2). Those functions measured for different $n$ will be scaled according to a unique set of $T_c$ and critical exponents to test if they can be collapsed. Specifically, $T$ needs to be scaled by $L^{1/\nu}(T - T_c)/T_c$, where $L = \sqrt[d]{n}$ and $d$ is the dimensionality of the system. $\chi$, $C$, and $M$ need to be scaled by $L^{-\alpha/\nu}$, $L^{\beta/\nu}$ and $L^{-\gamma/\nu}$, respectively. Achieving such a collapse implies that, at the thermodynamic limit, the system has a critical point at $T_c$, which is characterized by the divergence of $\chi$ and $C$ and the phase transition of $M$. To illustrate this, we consider the collapse of $\chi$, which implies that, at $T_c$, the scaled quantity of $\chi$, i.e., $L^{-\gamma/\nu}\chi$, is a constant. When $n \to \infty$, $L^{-\gamma/\nu} = n^{-\gamma/\nu d} \to 0$ because $\gamma/\nu d > 0$ (see below). Therefore, a finite product of $L^{-\gamma/\nu}$ and $\chi$ implies $\chi \to \infty$. We find an excellent collapse up to $n = 20$ (Fig. 2). Importantly, the values of $T_c$ estimated by the FSS method are close to 1 (Fig. 3, purple; Supplementary Table 1), suggesting that ensembles of neuronal avalanches are organized at the vicinity of a thermodynamic critical point. In addition to $T_c$, FSS also estimates the critical exponents, including $\nu$, $\alpha$, $\beta$ and $\gamma$. They characterize how $\chi$, $C$ and $M$ change as a function of $T$ at the thermodynamic limit. We find that $\nu \approx (0.8\text{-}0.9)/d$, $\alpha \approx 0.7$, $\beta$ close to 0 and $\gamma$ close to 1. These results are consistent across the datasets obtained from two monkeys (Fig. 3, purple; Supplementary Table 1).

**Avalanche dynamics at the macroscopic Scale**

Seeking to extrapolate from these results, we apply the FSS analysis to neural dynamics manifested at the macroscopic scale— the whole human brain— measured by MEG. It was found recently that ongoing neuronal activity in human MEG reflects neuronal avalanches as reported in previous non-human studies using the LFPs [22] (Supplementary Fig. S4). Despite of the dramatically different spatial scales between the LFP and MEG signals from monkeys and humans (>10000-fold difference in recording areas), we found strikingly similar behavior for the activity measured across the entire human cortex when the control parameter, $T$, and system size, n, change (Fig. 4). Again, FSS analysis suggests that $T_c \approx 1$ for the macroscopic system (Fig. 3, blue; Supplementary Table 1). Importantly, very similar exponents are obtained for the whole human brain recorded with MEG and the results are consistent across different human subjects (Fig. 3, blue). Such similarity, in terms of both the scaling behavior, i.e., collapse of curves, and critical exponents, strongly suggests a universal organization that underlies neuronal interactions at various spatial scales.

**Validating the FSS method through a simple model**

Next, we investigate a simple and understandable model, and demonstrate that the FSS analysis is sensitive enough to distinguish critical from non-critical systems. To this end, we used DG models in which all elements are embedded in a ring configuration. Each element has a well defined "distance" to any other element (Fig. 5A). We set the covariance of hidden variables (Methods) $i$ and $j$, $\lambda_{ij}$, as a Gaussian function of the distance $r_{ij}$ between them: $\lambda_{ij} = \lambda_{max} \exp\left[-\frac{1}{2}\left(\frac{r_{ij}}{\omega}\right)^2\right]$, where $\lambda_{max}$ is the maximal covariance and $\omega$ is the standard deviation of the Gaussian function. If $\omega \to \infty$, it approaches the situation that all $\lambda_{ij}$ are the same, for which criticality is ensured [11]. Decreasing $\omega$ to 0, drives the system to an independent state (Fig. 5B).

We applied the FSS method to this system. To facilitate the analysis, system sizes were set to be $n$=6-10. In Fig. 5C-F, we plot the goodness of collapse, estimation of $T_c$, and critical exponents as a function of $\omega$. We found that for this model, the deviation from the critical states ($\omega = \infty$) is detectable for $\omega$<7~8. Given that all $r \leq 5$, we consider the sensitivity of the FSS for detecting deviations from criticality as satisfactory. We note that with increasing system sizes included in the analysis, even higher sensitivity will be achieved. We also compared these results with real data ($n$=6-10) and found that the actual results we obtained for cortical activities are very close to a true critical state (Fig. 5C-F), further supporting the previous results that neuronal avalanches represent a cortical state close to thermodynamic criticality.

**Correlation structure in neuronal avalanche dynamics**

The results based on this simple model also provide testable predictions for the empirical data. First, if we remove all correlations in activities between cortical sites, the critical behavior observed for the original data should be abolished. To test it, we used independent Poisson processes to generate nLFPs with the same empirically measured rate for each cortical site. $\chi$, $C$, and $M$ were then calculated as a function of $T$ and $n$ in the same way for the original data. As expected, all three quantities do not depend on system size and thus do not show any scaling behavior (Supplementary Fig. S5). The second and more important prediction is that the original data should contain long range correlations. In Fig. 6, we plot the correlation $G$, defined as $G_{ij} = \langle \sigma_i \sigma_j \rangle - \langle \sigma_i \rangle \langle \sigma_j \rangle$, as a function of the Euclidian distance $r$ between sites $i$ and $j$ in both linear and log-log coordinates. We found that the fluctuations of activity between very distant cortical sites are highly correlated (Fig. 6A, B). More precisely, the correlation drops approximately as a power-law with an exponent of -0.25 (Fig. 6C, D). The slow and approximate power-law decay in spatial correlations is consistent with our conclusion that the resting cortical activity is organized close to a thermodynamic critical point [7]. It is interesting that the data and the model with $\omega=\infty$ share the same set of critical exponents (Fig. 5E, F), despite their differences in correlation structure. Whereas $G$ is constant in the model (for $\omega = \infty$), it changes systematically as a function of $r$ in the data. Consequently, all patterns with the same size are equally probable in the model [11], whereas these probabilities can differ by up to 2 orders of magnitude in the data. Therefore, the fact that the model and the data share the same set of exponents is non-trivial, suggesting that they belong to the same universality class.

**Relation between the power-law size distribution and thermodynamic criticality**

Importantly, we note that the equilibrium critical behavioral revealed here is not implied by the power-law distributed avalanche sizes. This can be seen clearly by studying the probability $p_0$ of the quiescent state, i.e., all sites are inactive. This probability is not constrained by the power-law distribution in avalanche patterns (because it leads to divergence for a power-law), but nevertheless is important in order to obtain proper scaling and collapse using FSS. In the original data, $p_0$ decreases in a unique way with an increase in system size $n$ (Fig. 7). If $p_0$ is randomly changed with $n$, the functions cannot be collapsed despite the preservation of the power-law size distributions (Fig. 8A). On the other hand, estimated $T_c$ remains close to 1, if the actual $p_0$ is kept while randomizing probabilities for all other patterns, i.e. destroying the power-law size distribution (Fig. 8B). This makes sense because how $p_0$ changes as a function of $n$ parsimoniously reflects the underlying correlation structure of the system and therefore, largely determines how far the system is away from the criticality. This can be also demonstrated by the fact that in the model we described above, criticality ($\omega >> r_{max}$) is associated with different functions of $p_0(n)$, compared with non-critical conditions (Supplementary Fig. S6).

Although the power-law size distribution is neither sufficient nor necessary for the thermodynamic critical behavior revealed here, by testing a wide range of $T$, we found that $T$ that minimizes the distance from a power-law and the actual distribution is very closer to 1 (0.99 ± 0.03; mean ± SD across eight sub-groups from 2 monkeys for the best fitting power-law and 1.03± 0.10 for the power-law with slope -1.5; Supplementary Fig. S7), demonstrating that there is a unique "temperature" associated with the avalanche dynamics. Given the fact that the power-law size distribution is not necessarily associated with the thermodynamic criticality, our finding that cortical dynamics exhibit these two features simultaneously is intriguing.

## Discussion

Our results suggest that neuronal avalanches at both mesoscopic and macroscopic scales manifest a cortical state near thermodynamic criticality. The critical exponents found are highly consistent among different subjects and are reasonably consistent across the two different scales and species. Such results are reminiscent of the well-known fact that, near the critical state, emergent behaviors do not depend on the specific microscopic realization of a system and, therefore, a multitude of systems can be categorized into a small number of universality classes based on their critical exponents [5–8]. Our results thus suggest a general principle governing the collective behaviors of cortical activities at different spatial scales.

A recent study [43] reported that experimental data might falsely imply criticality due to 1) the limitation of finite sampling and 2) the bias introduced when choosing parameters to achieve best accuracy in the inferring procedure. However, neither aspect applies to the current study. The pair-wise correlation we observed for nLFPs that constitute neuronal avalanches are within the range of 0.2-0.6 (Pearson's *r*) and given our sample sizes, the margin of error is <0.05 (95% confidence interval). Therefore, our sample sizes were large enough to infer even lower or higher correlation strengths (indicating larger distances from the critical state, see [43]), if they actually existed in the system. This suggests that the proximity to a critical state is a true feature of the cortex. Furthermore, in the current analysis, no parameter for analyzing the data was chosen according to the criterion of inferring accuracy. Taken together, the current results are robust, in light of the known methodological biases.

One of the key steps in our analysis is to use the single histogram method to infer the system's behavior for different values of the control parameter *T*. This is a well-established method and has been widely applied to study various empirical systems and models at, or close to equilibrium [10,11,40]. Using the same method, Stephens et al. [40] recently found that the spatial pattern of natural images contains indications of criticality. Macke et al. [11] found that if a system exhibits higher-order interactions, its specific heat will diverge as long as the correlation does not decay as a function of the distance. In a study of spiking activities in salamander retina [10], it was found that the maximal heat capacity increases with system size and the corresponding $T$ ($T_{peak}$) approaches 1. This was suggested as evidence for criticality [10]. Heat capacity, though, is an extensive quantity and thus, an increase in heat capacity with increasing system size is difficult to interpret. It does not necessarily indicate an increase in specific i.e., normalized, heat capacity. Furthermore, without a sound extrapolation of $T_{peak}$ as $n \to \infty$, it is difficult to give an accurate estimation of $T_c$. In the current study, we took several steps to avoid such ambiguities. First, specific heat $C$ was analyzed directly. More importantly, we use FSS to estimate both $T_c$ and the critical exponents, providing a quantitative characterization of the system's behavior.

Interestingly, the critical exponents derived for the cortical activities are different from those that are commonly found in physics such as the Ising model, Heisenberg model or Spherical model[7]. Cortical activity has distinctive features, including a currently unknown dimensionality and a special structure of higher-order interactions [30], which may underlie its unique critical exponents. We also notice that the value of $β$ is close to zero, which in some cases indicates that the phase transition is a discontinuous one [44]. However, recently it was found that some continuous phase transitions have $β$ so close to zero that it is practically indistinguishable from a discontinuous one [45]. To further elucidate this issue, future work with approaches that can analyze much larger systems, i.e., larger $n$, would be needed to increase the precision in estimating $T_c$ and critical exponents.

Our current approach did not address the organization of activities within individual avalanches. It has been previously demonstrated that such activities can be effectively understood in the framework of a critical branching process [14,20,27,29,46]. That approach considers the spatiotemporal organization of events (nLFPs) that occur in an avalanche to be the result of balanced cascades and correctly predicts the power-law distribution in avalanche size with the exponent of -1.5. The critical branching process is a well-studied, non-equilibrium critical condition, which belongs to the universality class of directed percolation [47]. By collapsing the temporal dimension, we compressed the spatiotemporal pattern of neuronal cascades into spatial-only patterns and thus ignore the non-equilibrium cascading process in our present study. At the same time, we analyzed the ensemble of all cascades as a whole. Thus, our approach focuses on the organization of avalanche activities at a different level. In this regard, the current results provide a complementary view to better understand cortical dynamics, suggesting a more sophisticated, but highly organized, hierarchical structure. We propose that cortical dynamics are organized close to criticality from both the non-equilibrium, branching and the equilibrium thermodynamic perspective. The former is indicated by a power-law size distribution, whereas the latter is indicated by $T_c$ close to 1. Interestingly, a recent study found that the interactions among brain areas may pose the whole brain close to an equilibrium critical state [48]. Future studies to investigate how the brain can achieve both types of criticality, at different spatial as well as temporal scales hold great promise to uncover a more complete picture of cortical dynamics.

For the non-equilibrium critical state characterized by power-law probability distributions, theoretical as well as empirical studies have revealed functional advantages for neuronal information processing [14,26–29,49]. Moreover, deviation from this state has been suggested to be related to pathological conditions [50]. The equilibrium, thermodynamic criticality also has direct functional implications. From an information-theoretic point of view, the maximal specific heat, i.e., maximal variance of $\log(p_i)$, implies largest dynamic range for population coding [10,11]. This is also consistent with the finding that the dynamics of the brain reach highest signal complexity near the equilibrium criticality [48]. The maximal susceptibility has an even more straightforward interpretation: it means that cortical networks have obtained largest sensitivity to small perturbation. This may play an essential role in allowing the organism to be able to detect and respond to subtle environment changes. Such a high sensitivity of cortical networks has been demonstrated empirically for both spiking activity [12,13] and neuronal population activity reflected in the LFPs [14]. The current results provide new insights into these intriguing phenomena of cortical dynamics.

Finally, what could be the neuronal interpretation of the fictitious temperature $T$? In general, the change of $T$ drives the system through a phase transition. That is, from an ordered phase, characterized by low activity and strong coupling, to a disordered phase, characterized by high activity and weak coupling. Available physiological evidence suggests that similar changes simulated here by changing $T$ might occur in the brain. The transition of the overall cortical state from a low-activity, strong-coupling regime to a high-activity, weak-coupling regime has been documented in sleep/awake and inattentive/attentive transitions [15–19]. These observations suggest that there might be intrinsic neural mechanisms for adjusting the overall cortical state, roughly along the same dimension as changing $T$. It is well known that neuromodulators, such as acetylcholine (ACh), are involved in controlling such state changes [51-53]. The action of ACh, probably combined with other neuromodulators, may serve as a natural mechanism underlying the regulation of cortical state by controlling a parameter similar to *T*. Consistent with this hypothesis, studies have reported that applying ACh to neuronal cultures with neuronal avalanche dynamics drives the system towards a high-activity, weak-coupling regime [54,55]. Similar effects have also been observed for spiking activities *in vivo* at the visual cortices [56,57]. Future experiment combined with well controlled ACh manipulation [56,58] and monitoring of large population activities would be able to test such hypothesis in a quantitative manner and shed new light on how the emerged properties of neuronal populations are regulated in the brain.

*Acknowledgments*

The authors would like to thank Richard Saunders and Andy Mitz for help with monkey data collection and the MEG Core Facility of the NIMH for help with human data collection. The authors would also like to thank Jayanth R. Banavar, Amos Maritan, Rajarshi Roy, Mauro Copelli, Mikhail Anisimov, Didier Sornette and Woodrow Shew for helpful comments on an early version of the paper. This study was supported by the Intramural Research Program of the National Institute of Mental Health, NIH.

*Author Contributions*

Conceived and designed the experiments: SY HY OS DP. Performed the experiments: SY OS. Analyzed the data: SY HY OS. Contributed reagents/materials/analysis tools: HY. Wrote the paper: SY HY OS DP.

*Competing Interests*

The authors have declared that no competing interests exist.

# Figure Legends

**Figure 1. Identifying avalanche dynamics in LFP signals.** *A*, Lateral view of the macaque brain showing the position of the multi–electrode array (square, not to scale) in pre-motor (Monkey 1; blue) and prefrontal (Monkey 2; orange) cortices. *PS*, Principal Sulcus. *CS*, Central Sulcus. *B*, Example period of continuous LFP at a single electrode. Asterisks indicate peaks of negative deflections in the LFP (nLFPs) that pass the threshold (*Thr.*, broken line; - 2.5 SD). *C*, Identification of spatiotemporal nLFP clusters and corresponding spatial patterns. Left: nLFPs that occur in the same time bin or consecutive bins of length *Δt* define a spatiotemporal cluster, whose size is given by its number of nLFPs (two clusters of size 4 and 5 shown; gray area). Right: Patterns represent the spatial information of clusters only. *D-E*, Neuronal avalanche dynamics are identified when the sizes of activity cascades distribute according to a power-law with slope close of -1.5. Four distributions from the same original data set (solid line) using different areas (inset), i.e., number of electrodes (*n*), are superimposed. The power-law distributions vanish for shuffled data (broken lines). A theoretical power-law with slope of -1.5 is provided as guidance to the eye (gray, broken line). *D* is reprinted from [30].

**Figure 2. Critical behavior in susceptibility, specific heat and order parameter observed for neuronal avalanches at the mesoscopic level, i.e., recorded by LFPs.** Susceptibility (*A*), specific heat (*B*) and order parameter (*C*) plotted as a function of *T* for system size *n*=12-20 (color code). Left: Original non-scaled functions. Right: Corresponding collapse using FSS analysis. Scaled quantities plotted as a function of $t = (T - T_c)/T_c$, $L = \sqrt[d]{n}$, where *d* is the dimensionality of the system. Critical exponents: *α, β, γ* and *ν*. We note that the peaks for the scaled variables *χ* and *C* are not expected to be at the location of $L^{1/\nu}t=0$.

**Figure 3.** $T_c$ **and critical exponents α, β, γ and ν estimated using finite size scaling analysis in two monkeys and three human subjects.** Four (two) different 20-electrode/sensor sub-groups were analyzed for each monkey (human) dataset resulting in the sample size of 8 (6). Values are mean (center circle) ± s.d. (error bars omitted for s.d. smaller than center circle).

**Figure 4. Critical behavior in susceptibility, specific heat and order parameter observed for neuronal avalanches in the human brain at macroscopic level, i.e., recorded with MEG.** Susceptibility (***A***), specific heat (***B***) and order parameter (***C***) plotted as a function of *T* for system size *n* = 12-20 (color code). Left: Original non-scaled functions. Right: Corresponding collapse using FSS analysis. Scaled quantities plotted as functions of "reduced temperature", $t = (T - T_c)/T_c$, $L = \sqrt[d]{n}$, where *d* is the dimensionality of the system. Critical exponents: α, β, γ and ν.

**Figure 5. Validating the FSS method by a simple model.** ***A***, All elements are configured in a ring and the distance between any adjacent elements is 1. ***B***, the covariance of the hidden variables in the DG model, λ, is plotted as a function of the distance, *r*, that separates corresponding elements for different choices of the standard deviation of a Gaussian function, ω. ***C-F***, goodness of collapse, $T_c$ and critical exponents measured for various systems are plotted against ω (open circles). In all systems, $\lambda_{max}$ and mean event rate were set such that when ω=∞, the average covariance and the event rate match what we empirically observed for Monkey 1. Corresponding results obtained from actual data for Monkey 1 (averaged across four sub-groups) are shown for comparison (broken lines).

**Figure 6. Correlation function for avalanche activities.** Pair-wise correlation, *G*, of nLFP activities plotted against the physical distance between the corresponding recording sites. ***A-B***, linear coordinates. ***C-D***, Double-logarithmic coordinates. A power-law with slope of -1/4 is provided for reference (broken lines).

**Figure 7. Probability of the quiescent state changes as a function of the system size.** For 4 subgroups analyzed in monkey 1, probability of the quiescent state measured for the original data (blue) is plotted as a function of the systems size (from 1 to 20). Probability of the quiescent state measured for corresponding shuffled data (orange) is plotted for comparison. Shuffled data is obtained by randomizing the activity sequence for individual electrodes, which eliminates the correlation among different electrodes but preserves the probability of being active for all electrodes.

**Figure 8. Double dissociation between the scaling/collapse and the power-law size distribution.** *A*, Pattern probabilities of the original data were modified so that the probability for the quiescent state, $p_0$, was set randomly from a uniform distribution (0 , 1) while the probabilities for all other states were renormalized, i.e., $p_i=p_i/(1-p_0)$. Therefore, the power-law size distribution was preserved. Left: Specific heat, *C*, was plotted as a function of *T* for system size *n*=12 - 20 (color coded). Right: No collapse can be achieved. *B*, Pattern probabilities of the original data were modified so that the $p_0$ was unchanged while the assignments of $p_i$ among all other patterns were shuffled. Therefore, the power-law size distribution was abolished. Left: Susceptibility, $\chi$, was plotted as a function of *T* for system size *n*=12-20 (same color code as in A). FSS analysis estimated the $T_c \approx 1.1$. Right: collapse of functions.

**Figures**

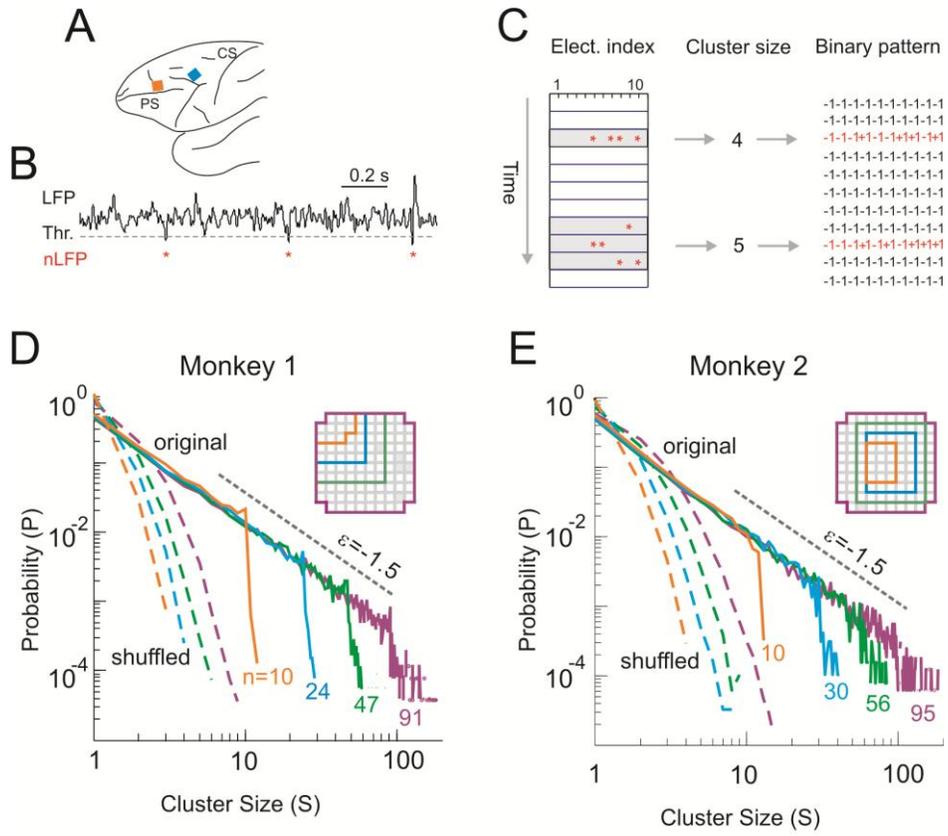

Fig. 1

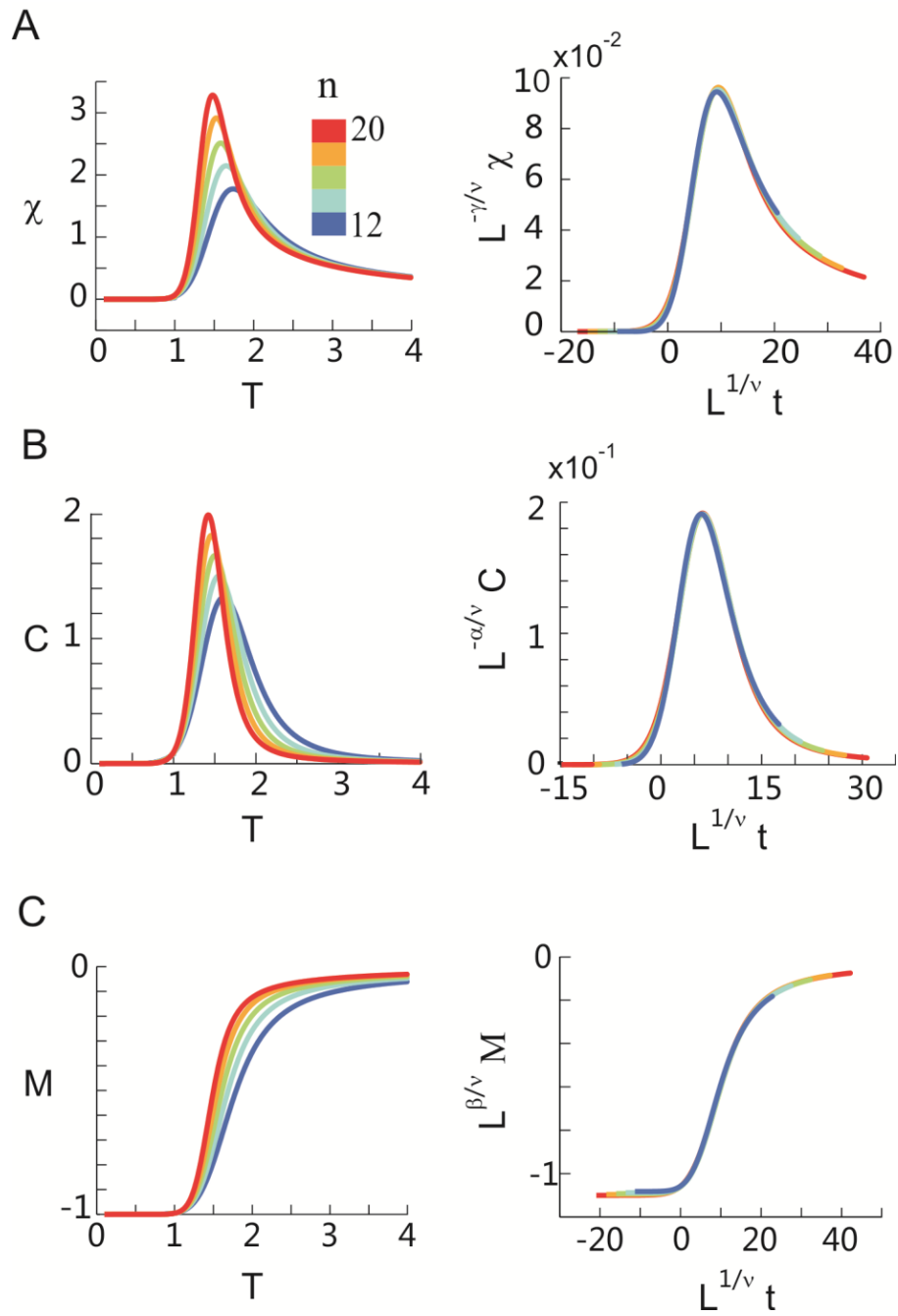

Fig. 2

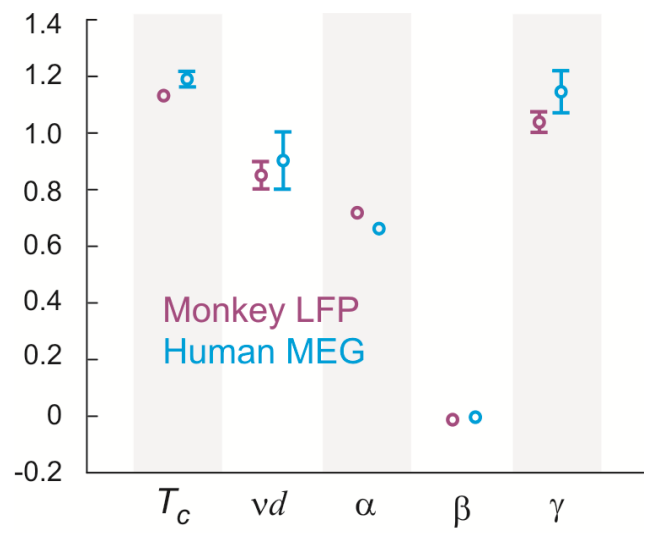

Fig. 3

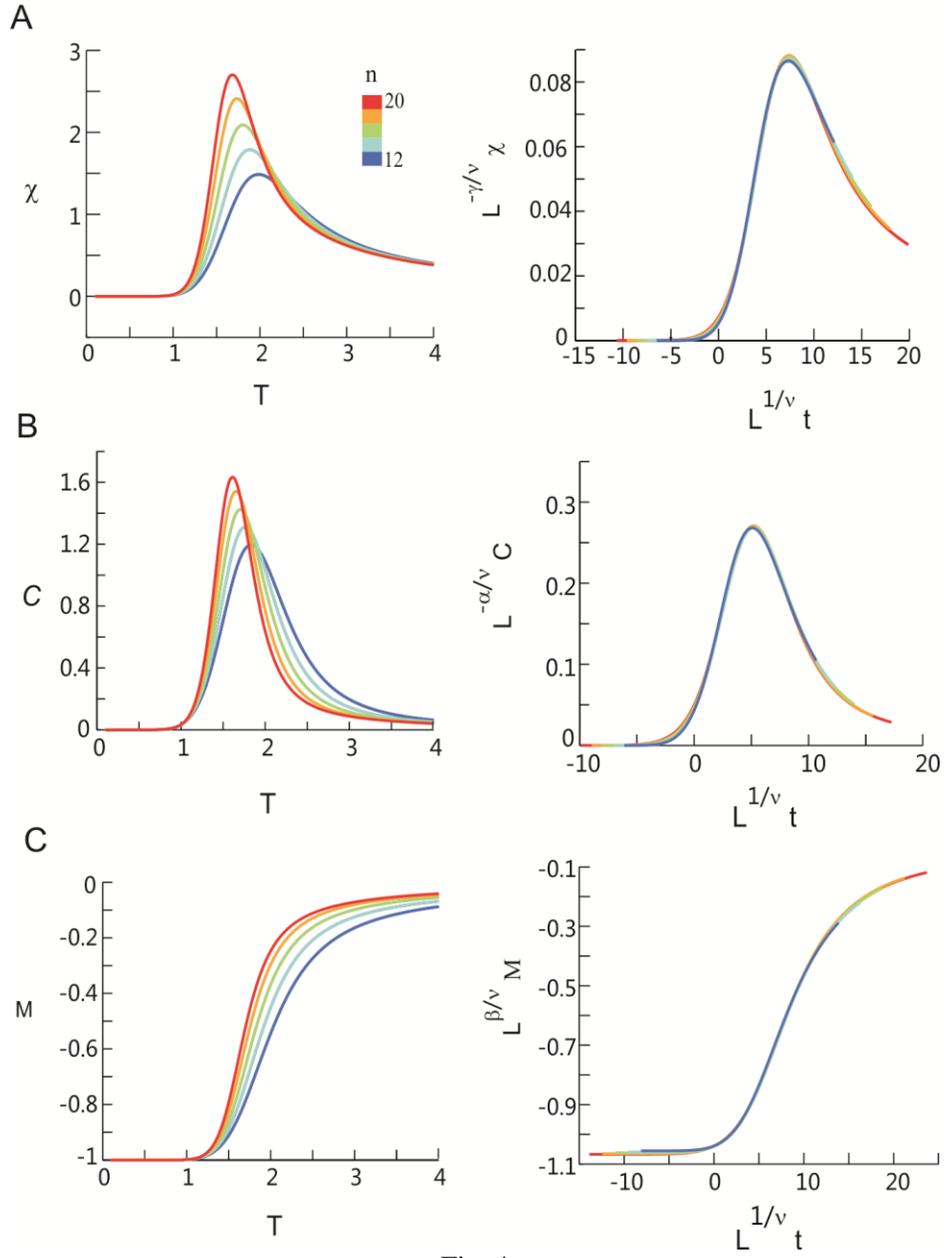

Fig. 4

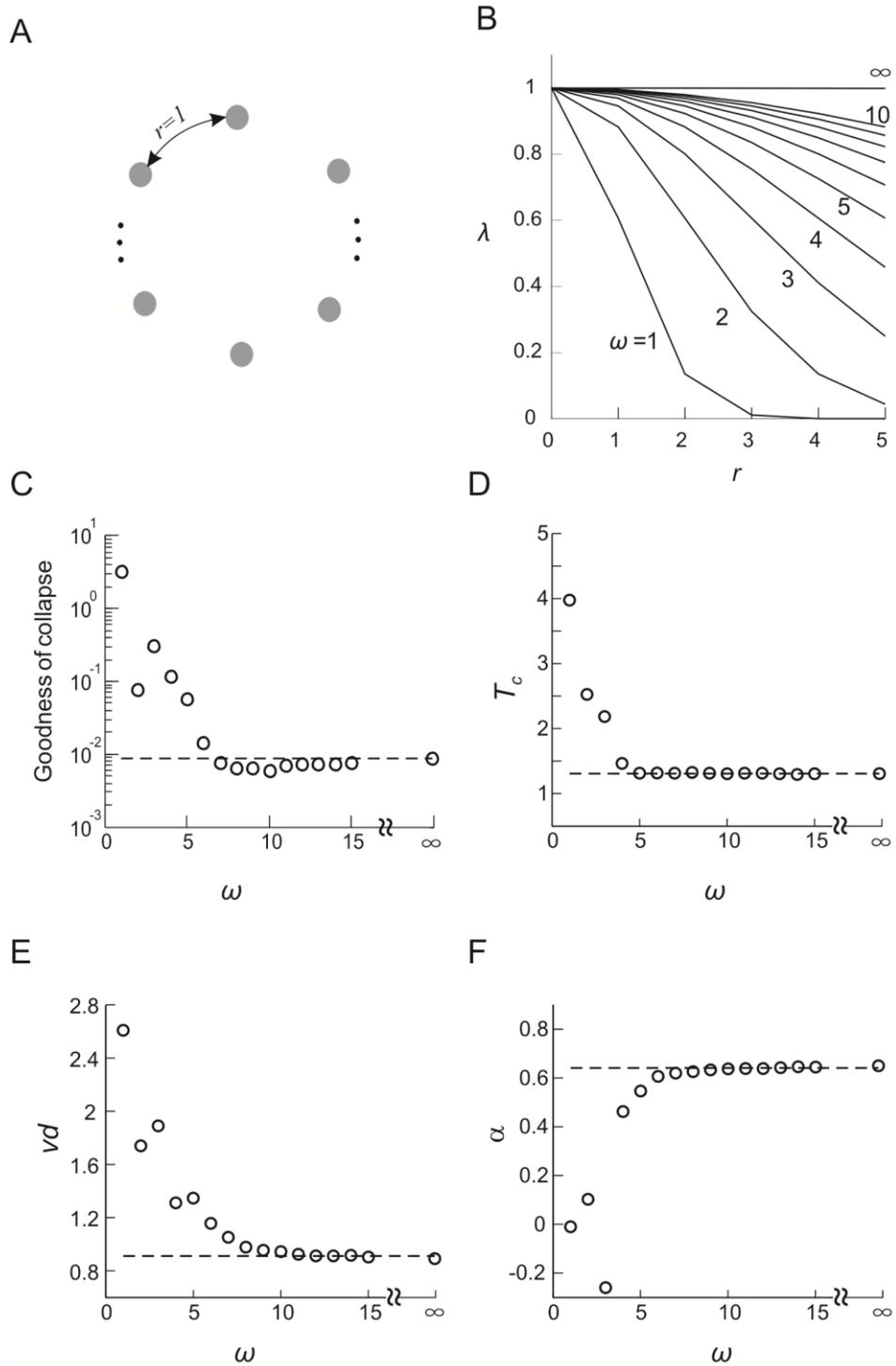

Fig. 5

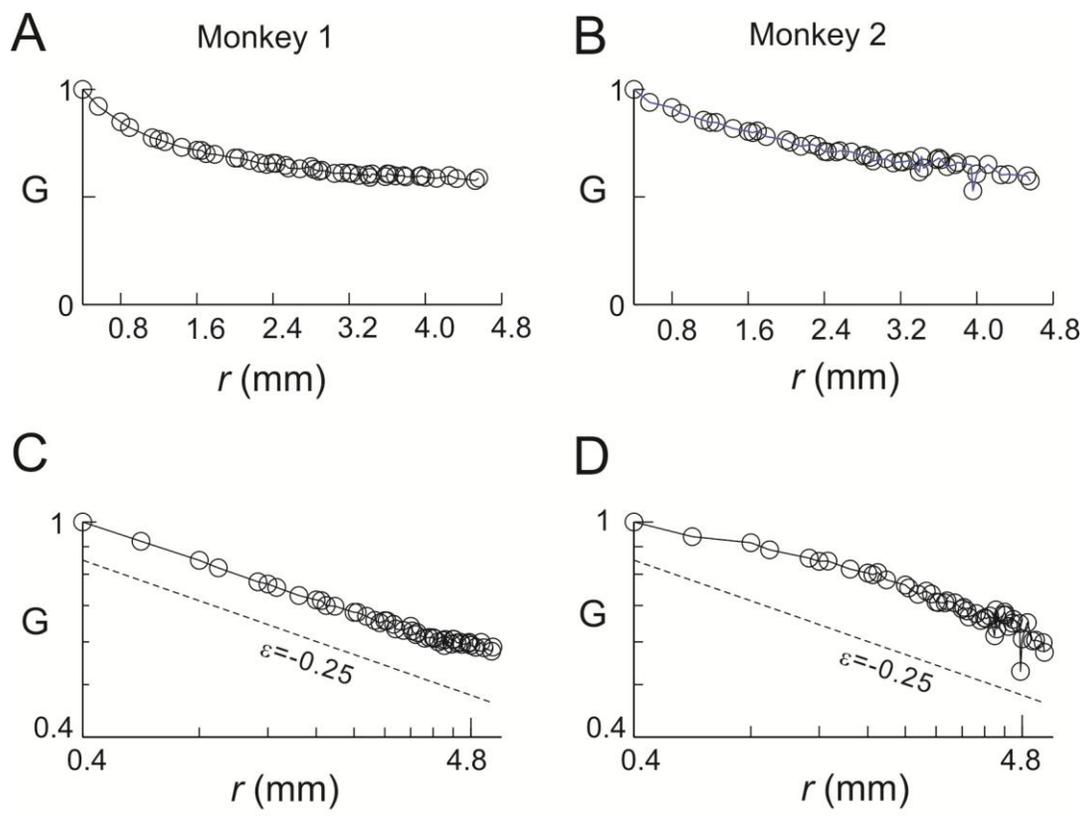

Fig. 6

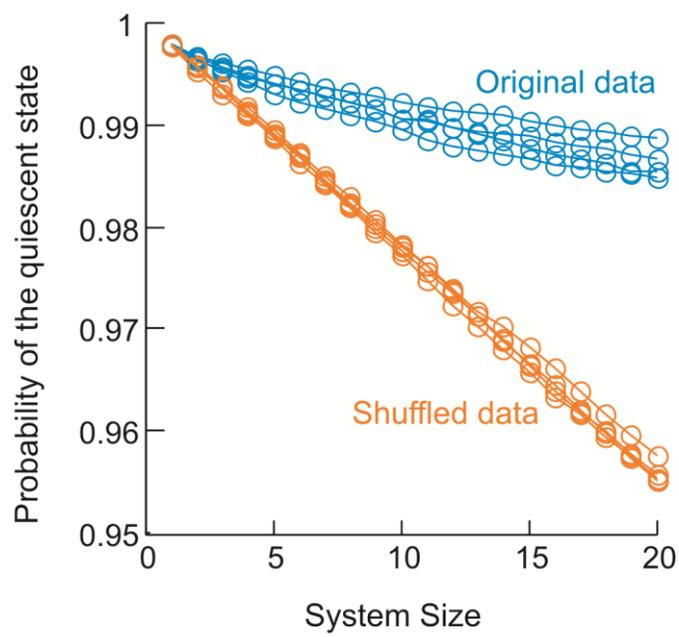

Fig. 7

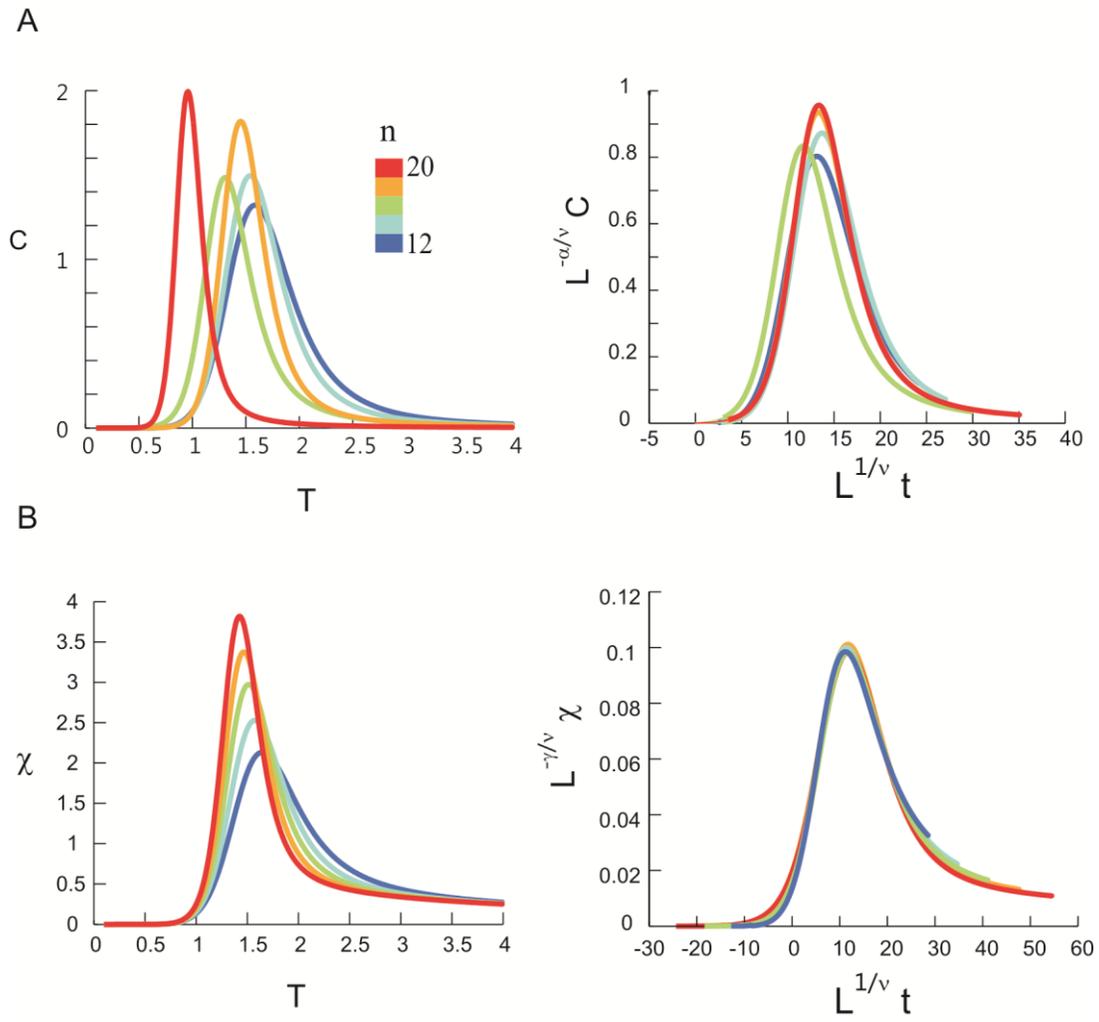

Fig. 8

# Supplementary Information

Index
I. Supplementary Methods
        A. Examining the assumptions about stationarity and equilibrium
        B. Analytical derivation of finite size scaling method
II. Supplementary Table
III. Supplementary Figures

# I. Supplementary Methods

### A. *Examining the assumptions about stationarity and equilibrium*

Thermodynamic equilibrium implies that the macroscopic properties of the system keep stable and do not change with time. As the distribution of avalanche sizes captures the essential feature of cortical dynamics [1–5], we examined the stability of this size distribution. In Fig. S3, we show that the avalanche size distribution, measured for 10 consecutive, equal-sized segments of recording, is stable across the whole recording period (30 min). To contrast this with the true equilibrium condition, we shuffled the original avalanche raster (i.e., randomized its sequence) and repeated the same analysis. The variability of the estimated power law exponent, $\varepsilon$, across all segments is small for both the original and shuffled datasets. F-test statistics also revealed no significant difference in the variance of $\varepsilon$ between the two conditions ($p=0.13$), suggesting a stable organization of the system over time.

Secondly, we demonstrate that the data satisfy two crucial criteria that will lead to equilibrium: 1) detailed balance (micro-reversibility) and 2) accessibility (ergodicity) [6]. Detailed balance is achieved in a system if the following relation holds: $p_i p_{i \to j} = p_j p_{i \to i}$, where $i$ and $j$ are possible states (configurations) of the system; $p_i$ is the probability of states $i$ and $p_{i \to j}$ is the transition probability from state $i$ to state $j$. For avalanche patterns defined by clustering a period of activity flanked by quiescent periods before and after it [see method section for details; see also [1]], it is clear that the detailed balance strictly holds for systems with arbitrary sizes. As in this condition, every transition from a quiescent state (i.e., all sites are inactive) to an active state (i.e., at least one of the sites is active) would be accompanied by a reverse of that transition. In other words, the system will satisfy $p_q p_{q \to i} = p_i p_{i \to q}$, where $q$ is the quiescent state and $i$ is any active state. Such a feature, combined with the fact that all $p_{i \to j} = 0$ when $i$ and $j$ are both active states, ensures the detailed balance.

To study whether the detailed balance still holds when we release these constraints set by the rules that identify avalanches, we examined the relation between $p_i p_{i \to j} = p_j p_{j \to i}$ in the data with quiescent periods removed. In such case, both constraints, i.e., the symmetrical transition from a quiescent state to an active state and zero transition probability between active states, are removed. In Fig. S4, we plotted the measured $p_i p_{i \to j}$ against $p_j p_{j \to i}$ for systems with different sizes ($n$=2-5). Overall, the data points are fairly close to the identical line, suggesting the fulfillment of the equality. For comparison, we constructed a shuffled data set, in which the sequence of avalanche patterns was randomized. For this shuffled data set, any possible temporal dependency was removed so it is in a truly equilibrium state and, therefore, fulfills the detailed balance. The same analysis was then performed for the shuffled data and we found that the results are similar to those from the original data, indicating that the deviation from the identical line for the original data is largely due to finite sampling, and not due to a violation of the detailed balance. To quantify this effect, we computed the ratio $r = D_{data} / D_{shuffled}$, where

$$D = \frac{2 | p_i p_{i \to j} - p_j p_{j \to i} |}{p_i p_{i \to j} + p_j p_{j \to i}} \quad (1)$$

We found that for $n$>2, this ratio is very close to one (Mann–Whitney U test, $p$>0.05), indicating that the violation to the detailed balance is sufficiently small so it is not detectable within the current recording length. Due to the lack of sufficient data, the direct check of detailed balance cannot be performed for larger systems (n >> 5). However, with the results we obtained for $n$=2-5, and given the fact that with the increase of system size, exponentially more samples would be needed to detect the same level of violation, it is clear that the detailed balance among the active states, i.e. avalanche patterns, should be a good approximation for even larger systems.

Regarding the accessibility/ergodicity assumption, it requires that from any given state, the system should be able to evolve (after a sufficiently long time) to any other state. Although the direct test for ergodicity is not possible due to limited length of the recording, the power-law distribution in avalanche sizes provides strong empirical evidence to support it. Such a heavy-tailed distribution indicates that even large systems can visit configurations that cover all possible avalanche sizes.

Taken together, various empirical tests strongly suggest that the stationarity and even the equilibrium assumption can be considered a reasonable first approximation for our data.

### B. Analytical derivation of finite size scaling method

For readers who are not familiar with the finite size scaling, we illustrate the method using susceptibility $\chi$ as an example. In the vicinity of the critical temperature $T_c$, $\chi$ can be expressed as a function of correlation length $\xi$.

$$\chi = \xi^{\gamma/\nu} \tag{2}$$

In finite size system, correlation length $\xi$ is comparable to system size $L$, and therefore has a cut off. Consequently, $\chi$ also has a cut off. If we use $\xi$ to represent the correlation length at the thermodynamic limit, then the cut off takes place when $\xi > L$. Then, we can rewrite $\chi$ as

$$\chi = \xi^{\gamma/\nu} \chi_0(L/\xi), \tag{3}$$

which satisfies the conditions above. Then define

$$\chi_0(x) \sim x^{\gamma/\nu}, \quad \text{for } x < 1$$

$$\chi_0(x) \sim c, \quad \text{otherwise, where } c \text{ is a constant.} \tag{4}$$

Therefore, when the system size is finite,

$$\chi_L = \xi^{\gamma/\nu}(L/\xi)^{\gamma/\nu} = L^{\gamma/\nu}, \tag{5}$$

And the correlation length is comparable to the system size. Otherwise, when the system size is infinite, the correlation length is actually $\xi$,

$$\chi = c\xi^{\gamma/\nu}. \tag{6}$$

Now we can rewrite the equation in order to remove $\xi$, because we do not know its exact value, and also to introduce a dimensionless function $\bar{\chi}(x)$, which will be the scaling function for $\chi_L$

$$\begin{aligned}\chi &= \xi^{\gamma/\nu} \chi_0(L|t|^\nu) \\ &= \xi^{\gamma/\nu} \chi_0[(L^{1/\nu}|t|)^\nu] \\ &= |t|^{-\gamma} \chi_0[(L^{1/\nu}|t|)^\nu] \\ &= L^{\gamma/\nu} L^{-\gamma/\nu} |t|^{-\gamma} \chi_0[(L^{1/\nu}|t|)^\nu] \\ &= L^{\gamma/\nu}(L^{1/\nu}|t|)^{-\gamma} \chi_0[(L^{1/\nu}|t|)^\nu]\end{aligned} \tag{7}$$

Set $x = L^{1/\nu}|t|$, which will be the scaling variable

$$\chi = L^{\gamma/\nu} x^{-\gamma} \chi_0(x^\nu). \tag{8}$$

Define scaling function $\bar{\chi}(x) = x^{-\gamma} \chi_0(x^\nu)$, then

$$\chi = L^{\gamma/\nu} \bar{\chi}(L^{1/\nu} |t|). \tag{9}$$

Note when $\xi \sim L$,

$$\begin{aligned}
\bar{\chi}(x) &= x^{-\gamma} \chi_0(x^\nu) \\
&= (L^{1/\nu} |t|)^{-\gamma} \chi_0[(L^{1/\nu} |t|)^\nu] \\
&= L^{-\gamma/\nu} |t|^{-\gamma} \chi_0(L |t|^\nu) \\
&= L^{-\gamma/\nu} |t|^{-\gamma} c(L |t|^\nu)^{\gamma/\nu} \\
&= L^{-\gamma/\nu} |t|^{-\gamma} c L^{\gamma/\nu} |t|^\gamma \\
&= c
\end{aligned} \tag{10}$$

Thus, the scaling function is a constant and independent of the system size.

The scaling function also can be written as

$$\begin{aligned}
\bar{\chi}(L^{1/\nu} |t|) &= L^{-\gamma/\nu} |t|^{-\gamma} c(L |t|^\nu)^{\gamma/\nu} \\
&= L^{-\gamma/\nu} (|t|^{-\nu})^{\gamma/\nu} c(\frac{L}{|t|^{-\nu}})^{\gamma/\nu}
\end{aligned} \tag{11}$$

Recall $\xi \sim |t|^{-\nu}$, so we have

$$\begin{aligned}
\bar{\chi}(L^{1/\nu} |t|) &= L^{-\gamma/\nu} \xi^{\gamma/\nu} c(\frac{L}{\xi})^{\gamma/\nu}
\end{aligned} \tag{12}$$

Also recall, when system size is finite,

$$\chi_L = \xi^{\gamma/\nu} (L/\xi)^{\gamma/\nu}, \tag{13}$$

So

$$\bar{\chi}(L^{1/\nu} |t|) = L^{-\gamma/\nu} \chi_L = c. \tag{14}$$

From Eq. 19, we can measure $\chi_L(t)$ for various system sizes $L$ in a temperature range close to $T_c$, and rescale $\chi_L(t)$ by $L^{-\gamma/\nu}$ for each $L$ to obtain the scaling function $\bar{\chi}(L^{1/\nu} |t|)$, with $L^{1/\nu} |t|$ as the scaling variable. If we choose the correct $T_c$, $\nu$ and $\gamma$, the scaling functions for different system sizes will fall on the same curve.

# II. Supplementary Table

**Table S1** Critical temperature $T_c$ and critical exponents $vd$, $\alpha$, $\beta$ and $\gamma$ estimated using finite size scaling analysis (FSS) for eight 20-eletrode sub-groups in two monkeys (M1, M2) and six 20-sensor sub-groups in three human subjects (H1-H3). Arguments in brackets indicate that $T_c$ and $vd$ were estimated by applying FSS to susceptibility $\chi$, specific heat $C$ and order parameter $M$, respectively.

| Subject | Group | $T_c\,(\chi)$ | $vd\,(\chi)$ | $\gamma$ | $T_c\,(C)$ | $vd\,(C)$ | $\alpha$ | $T_c\,(M)$ | $vd\,(M)$ | $\beta$ |
|---|---|---|---|---|---|---|---|---|---|---|
| M1 | A | 1.13 | 0.88 | 1.04 | 1.15 | 0.92 | 0.72 | 1.16 | 0.84 | -0.028 |
|  | B | 1.12 | 0.86 | 1.00 | 1.14 | 0.90 | 0.72 | 1.14 | 0.84 | -0.021 |
|  | C | 1.12 | 0.86 | 0.98 | 1.14 | 0.88 | 0.72 | 1.13 | 0.84 | 0.001 |
|  | D | 1.12 | 0.86 | 1.02 | 1.15 | 0.88 | 0.73 | 1.16 | 0.80 | -0.03 |
| M2 | A | 1.10 | 0.82 | 1.05 | 1.14 | 0.84 | 0.71 | 1.16 | 0.76 | -0.03 |
|  | B | 1.11 | 0.90 | 1.10 | 1.13 | 0.96 | 0.71 | 1.13 | 0.84 | 0.001 |
|  | C | 1.10 | 0.84 | 1.06 | 1.14 | 0.84 | 0.71 | 1.12 | 0.78 | 0.001 |
|  | D | 1.11 | 0.82 | 1.05 | 1.15 | 0.86 | 0.72 | 1.13 | 0.78 | 0.000 |
| H1 | A | 1.16 | 0.84 | 1.20 | 1.22 | 0.86 | 0.67 | 1.20 | 0.74 | 0.0006 |
|  | B | 1.20 | 1.04 | 1.18 | 1.23 | 1.06 | 0.64 | 1.24 | 0.96 | -0.02 |
| H2 | A | 1.17 | 0.82 | 1.21 | 1.22 | 0.84 | 0.68 | 1.20 | 0.74 | -0.0007 |
|  | B | 1.18 | 0.98 | 1.17 | 1.22 | 1.00 | 0.66 | 1.20 | 0.92 | -0.0003 |
| H3 | A | 1.14 | 0.82 | 1.09 | 1.17 | 0.86 | 0.67 | 1.16 | 0.78 | 0.0007 |
|  | B | 1.18 | 0.98 | 1.02 | 1.20 | 1.00 | 0.65 | 1.17 | 0.98 | 0.0001 |

# III. Supplementary Figures

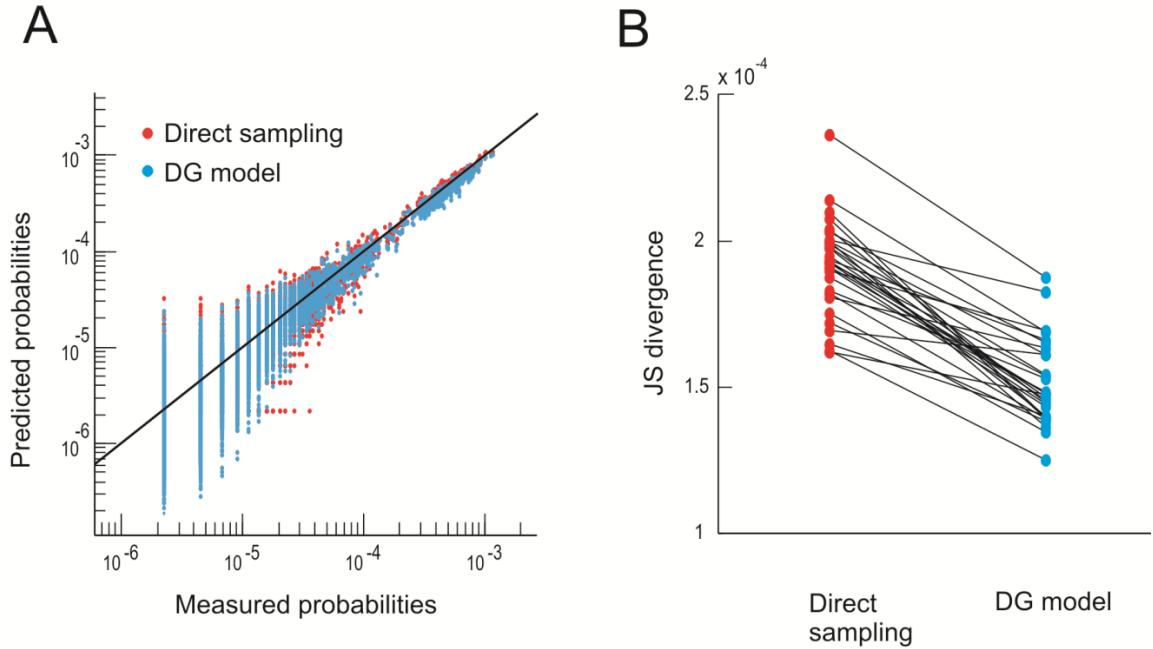

**Figure S1.** The DG model predicts state probability more accurately than direct sampling. **A,** Observed probability $p_i$ (thirty 10–electrode sub-groups) is plotted against the prediction made by direct sampling and the DG model. Solid line indicates equality. The comparison is based on 2-fold cross-validation[7]. **B,** JS divergence [7] between the observed and predicted probabilities of spatial avalanche patterns for the same thirty 10–electrode groups shown in (A). Linked dots are the results obtained by direct sampling and the DG model for the same group. The DG model has significantly smaller JS divergence (21% reduction, $p<10^{-5}$, paired–sample signed rank test).

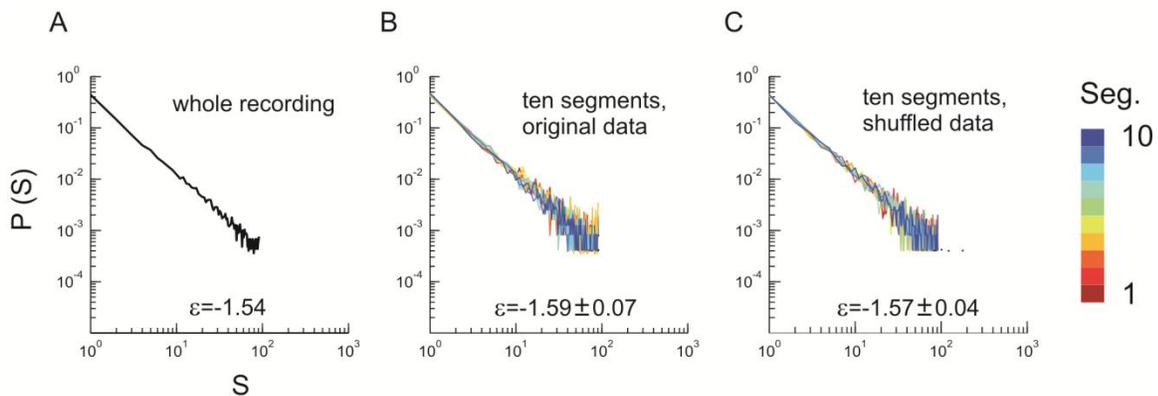

**Figure S2.** Stability of the power law size distribution during the recording. **A**, Avalanche pattern size distribution of the whole recording (30 min) plotted in a double-logarithmic scale. $\varepsilon$, exponent of the best fitting power law to the distribution. Avalanche pattern was indentified based on the activities recorded in the whole array (91 channels, Monkey 1). **B**, The whole dataset as analyzed in (A) was split into ten consecutive segments, each of which lasted for 3 min. Avalanche pattern size distributions were calculated for individual segments and plotted (color coded). *C*, The original dataset as analyzed in (A) was shuffled in time (i.e., the sequence of activities was randomized) to eliminate the temporal dependency. Then this shuffled dataset was split into ten consecutive, equal-sized segments. Avalanche pattern size distributions were calculated for individual segments and plotted (color coded). In (**B**) and (**C**), $\varepsilon$ is represented as mean ± s.d. (across all segments).

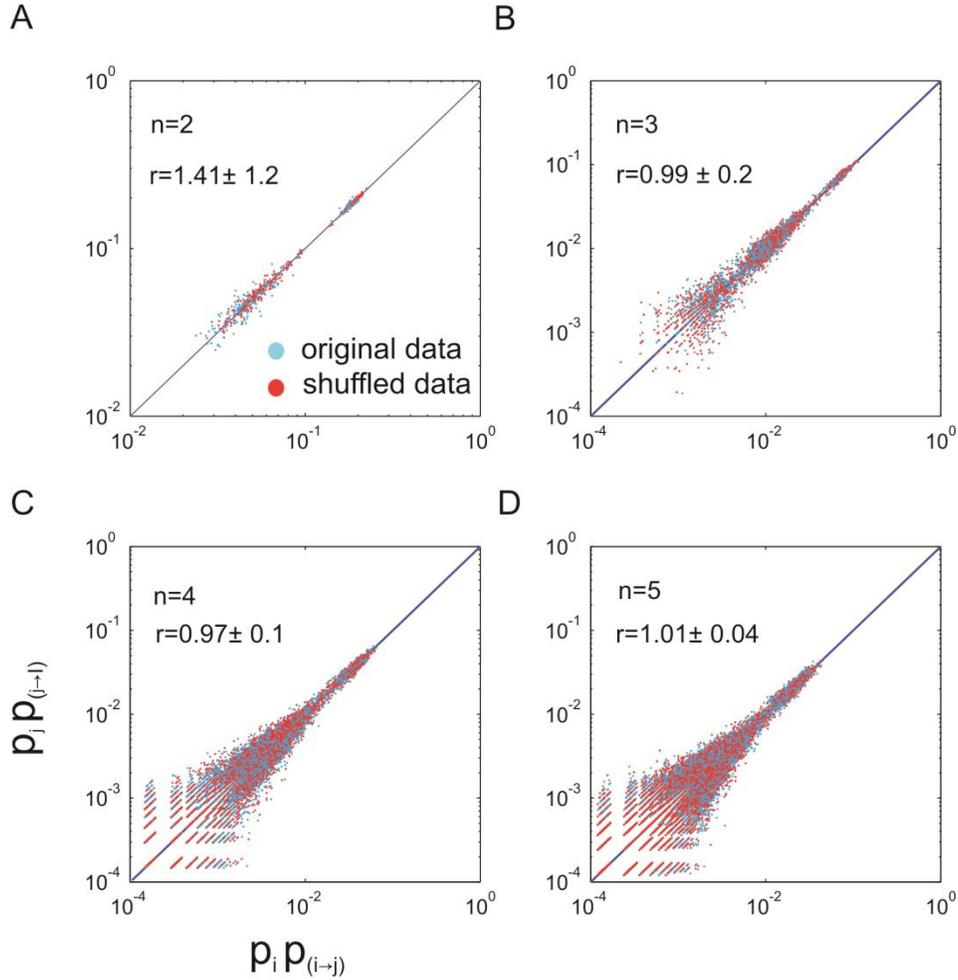

**Figure S3.** Detailed balance approximately holds for the data with quiescent periods removed. For differently sized systems (*n*=2-5), empirically measured $p_i p_{i \to j}$ is plotted against $p_j p_{i \to i}$ for both the original data (*blue*) and shuffled data (*red*). For every size, 100 different systems (i.e., different combinations of electrodes) were analyzed. The solid lines represent equality. *r* is a measure of the distance from the equality, relative to that of the shuffled data (see supporting text A for details). It is represented as mean ± s.d. (across 100 systems). A-D, system size equals 2, 3, 4 and 5, respectively.

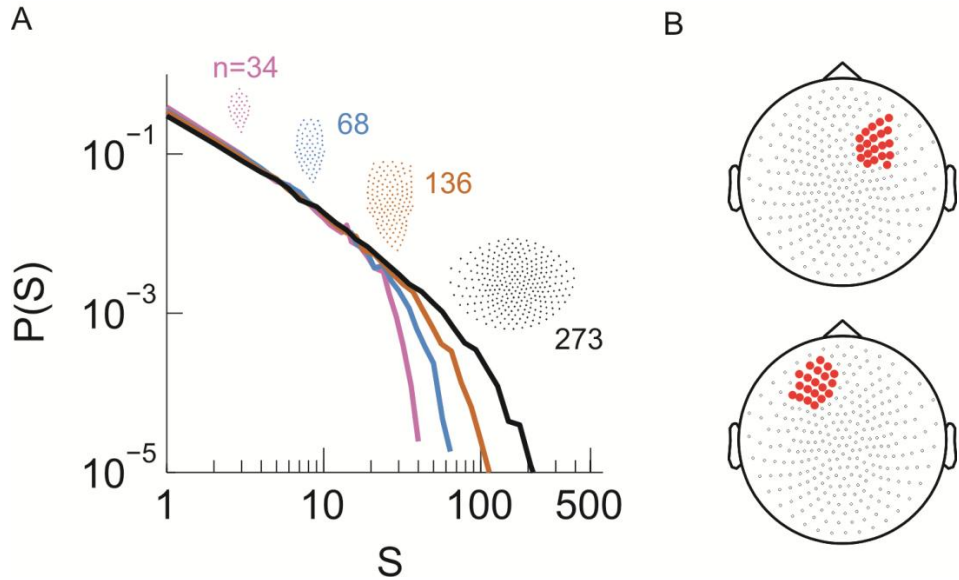

**Figure S4.** Power law size distribution in neuronal avalanches recorded with MEG for the human brain at resting state. **a**, Neuronal avalanche dynamics are identified when the sizes (S) of all clusters distribute according to a power law with slope close of −1.5 (the results for subject 2 are shown here). Four distributions from the same original data set using different areas (*insets*), i.e., number of MEG sensors (*n*), are superimposed. **b**, The whole array of sensors (*grey dots*) and two sub-groups of sensors that were used for finite-scaling analysis (*red dots*) are illustrated. Top, sub-group A; bottom, sub-group B. The sub-groups were identical across all three subjects.

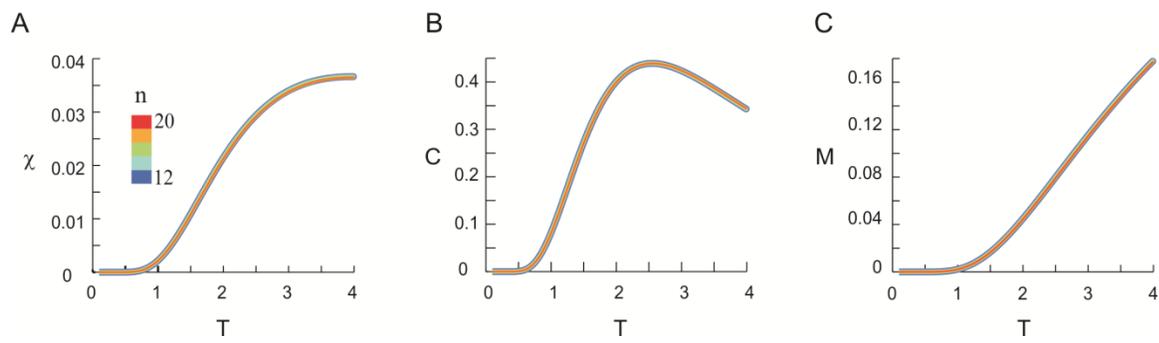

**Figure S5.** Control analysis (independent model). Original data was the same as shown in Fig. 2. At $T=1$, we calculated the individual pattern probabilities based on independent Poisson processes to generate nLFPs with the same empirically measured rate for each cortical site. Using the same method applied to original data, we calculate $\chi$, $C$ and $M$ as functions of $T$. In contrast to the original data, the curves for systems of different sizes are almost identical for $\chi$ (A), $C$ (B) and $M$ (C). For visual clarity, curves with different sizes have different widths.

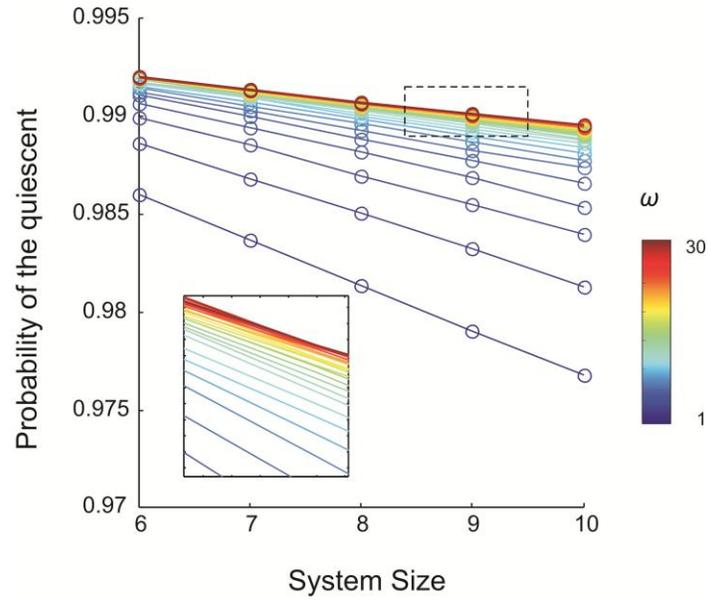

**Figure S6.** Probability of the quiescent state changes as a function of the system size in the model. Probability of the quiescent state measured for the model is plotted as a function of the systems size (from 6 to 10) with different $\omega$ (color coded). Inset: Zoon-in view of the area indicated by the rectangle (broken line).

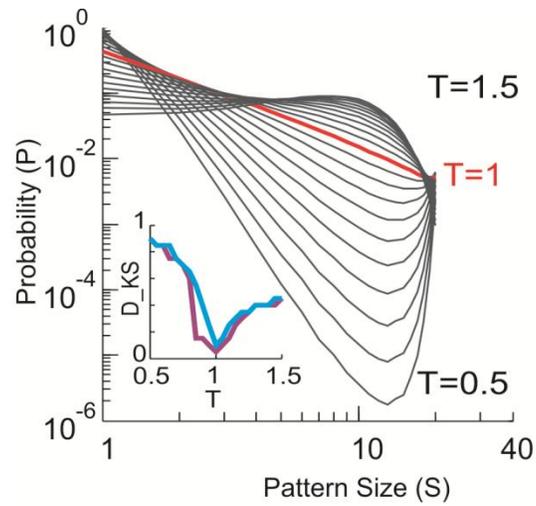

**Figure S7.** Size distributions of avalanche patterns computed for one 20-electrode sub-group (taken from data set in Fig. 1C) for different $T$ and plotted in double logarithmic coordinates. $T$ changes from 0.5 to 1.5 with a step of 0.1. Distribution at $T=1$ is marked by red. Inset: Kolmogorov-Smirnov distance ($D_{KS}$, a goodness-of-fit measure) between the actual pattern size distributions and best fitting power law (purple) or power law with slope -1.5 (blue) is minimized for $T \approx 1$.